\documentclass[letterpaper,aps,prl,twocolumn,tightenlines,nofootinbib,showkeys,superscriptaddress,longbibliography,preprintnumbers]{revtex4-2}

\pdfoutput=1
\usepackage{lipsum}
\usepackage[dvipsnames]{xcolor}
\usepackage{graphicx,epic,eepic,epsfig}
\usepackage[export]{adjustbox}
\usepackage{tikz,contour} 
\usepackage{feynmf}
\usetikzlibrary{shapes,arrows,positioning,automata,backgrounds,calc,er,patterns}
\usepackage[compat=1.1.0]{tikz-feynman}
\usepackage{tikz}
\usepackage{amsmath}
\usepackage{graphicx}
\usepackage{relsize}
\usepackage[left=1.8cm,right=1.8cm,top=1.8cm,bottom=1.8cm]{geometry}
\usepackage{XCharter}
\usepackage{mathptmx}
\usepackage[T1]{fontenc}
\usepackage{pifont}
\usepackage{amsfonts,amsmath,amssymb,amsbsy,bm,mathtools,slashed,latexsym,esvect,simpler-wick}

\usepackage{float,multirow,array,cancel,dcolumn}
\usepackage{enumitem}
\usepackage{hyperref}
\hypersetup{colorlinks=true,citecolor=red,linkcolor=NavyBlue,urlcolor=NavyBlue}
\usepackage[caption=false,labelformat=simple]{subfig}
\usepackage{lipsum}

\usepackage{comment}
\usepackage{shorthand}
\usepackage{natbib}

\renewcommand{\section}[1]{%
  \par\vspace{1ex} % Small space before, matching a paragraph break
  \noindent\textit{#1}.--- % Italics followed by em-dash
  \ignorespaces
}
\begin{document}

\preprint{UUITP-09/26}
\relscale{1.05}

\title{Testing varying coupling constants through multi-Higgs production at the LHC}

\author{Ulf Danielsson}
\email{ulf.danielsson@physics.uu.se}
\affiliation{Department of Physics and Astronomy, Uppsala University, Box 524, SE-751 20 Uppsala, Sweden}

\author{Rikard Enberg}
\email{rikard.enberg@physics.uu.se}
\affiliation{Department of Physics and Astronomy, Uppsala University, Box 524, SE-751 20 Uppsala, Sweden}

\author{Gunnar Ingelman}
\email{gunnar.ingelman@physics.uu.se}
\affiliation{Department of Physics and Astronomy, Uppsala University, Box 524, SE-751 20 Uppsala, Sweden}

\author{Soumyadip Kundu}
\email{soumyadip23@iisertvm.ac.in}
\affiliation{Indian Institute of Science Education and Research Thiruvananthapuram, Vithura, Kerala, 695 551, India}

\author{Tanumoy Mandal}
\email{tanumoy@iisertvm.ac.in}
\affiliation{Indian Institute of Science Education and Research Thiruvananthapuram, Vithura, Kerala, 695 551, India}

\author{Subhadip Mitra}
\email{subhadip.mitra@iiit.ac.in}
\affiliation{Center for Computational Natural Sciences and Bioinformatics, International Institute of Information Technology, Hyderabad 500 032, India}
\affiliation{Center for Quantum Science and Technology, International Institute of Information Technology, Hyderabad 500 032, India}

\begin{abstract}
\noindent
We propose the One Scalar Theory (1ST), a minimalist framework where a single real singlet scalar field mediates the dynamical generation of the Higgs self-coupling and the top Yukawa coupling. Unlike generic portal models, the 1ST removes parametric freedom by locking production and decay modes to a single fundamental scale $\Lambda_0$, rendering the framework highly predictive with unique experimental signals. We demonstrate that the collider phenomenology is partitioned by the $2m_t$ kinematic threshold into di-Higgs and di-top resonance regimes. By recasting current ATLAS data, we set lower bounds on $\Lambda_0$ at the TeV scale and show that the High-Luminosity LHC will probe this scale up to $4$~TeV, providing a definitive test for the dynamical origin of the electroweak sector.
\end{abstract}

\maketitle 
%%%%%%%%%%%%%%%%%%%%%%%%%%%%%%%%%%%%%%%%%%%%%%%%%%%%%%%%%%%%%%%%

%%%%%%%%%%%%%%%%%%%%%%%%%%%%%%%%%%%%%%%%%%%%%%%%%%%%%%%%%%%%%%%%
\section{Introduction}
Couplings and constants in a fundamental theory should emerge from the underlying dynamics, leaving no additional free parameters~\cite{Bekenstein:1982eu}. In string theory, for instance, couplings are determined by the vacuum expectation values (VEVs) of scalar moduli fields. These values set the fundamental constants of any low-energy effective theory realisation. If the Standard Model (SM) is realised in this manner, these scalar fields determine the parameters currently assumed to be free. Exploring the phenomenological consequences of scalar fields that dynamically generate these couplings offers a direct probe into this underlying structure~\cite{Danielsson:2016nyy,Danielsson:2019ftq}. Such variations can significantly alter cosmological and particle physics landscapes. Dynamically generated Yukawa couplings may trigger a strongly first-order electroweak phase transition, necessary for baryogenesis~\cite{Baldes:2016rqn,Braconi:2018gxo}. Similarly, a scalar field that controls the strong interaction strength can lead to early QCD confinement that allows QCD to remain confined at high temperatures~\cite{Ipek:2018lhm}, or it may make the QCD phase transition into a first-order transition that generates gravitational waves~\cite{Chatrchyan:2025wop}.While these effects have deep cosmological implications, their manifestation at the TeV scale provides a concrete target for collider experiments.

In this letter, we examine the Higgs self-coupling ($\lambda_h$) and the top Yukawa coupling ($\lambda_t$) because their interplay is central to the hierarchy problem, and the mechanism that stabilises the electroweak scale remains unknown. These two couplings are among the least constrained in the SM and govern di-Higgs production. Moreover, in the SM, the gluon-gluon fusion channel for di-Higgs production involves a delicate cancellation between the top-quark box and triangle diagrams. The resulting tiny cross-section makes this process a sensitive probe for new physics.

We propose a minimalist framework: the One Scalar Theory (1ST), where a single real singlet scalar field, $S_0$, governs the variations of both $\lambda_h$ and $\lambda_t$. In general, there could be two different scalars associated with the variations of two couplings. From the top-down perspective, there is no reason to expect that these scalars are aligned with the mass eigenstates. Furthermore, without fine-tuning, one would expect the lightest scalar to give the dominating contribution at low energies. This scalar would then couple to the Higgs as well as the top with roughly the same strength, suggesting that the 1ST model may serve as a good approximation to what can be expected from experiments.
%%%%%%%%%%%%%%%%%%%%%%%%%%%%%%%%%%%%%%%%%%%%%%%%%%%%%%%%%%%%%%%%
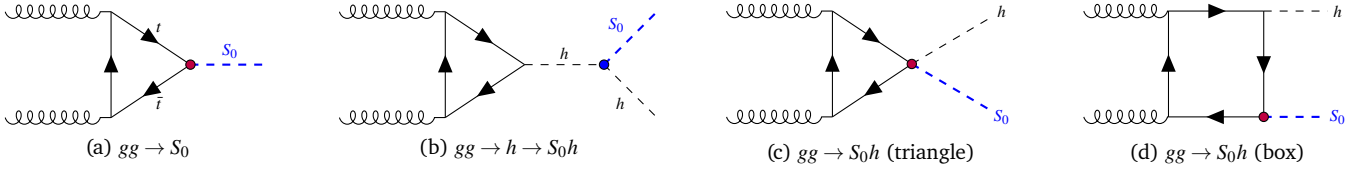
\begin{figure*}
    \centering
    % Define a global scale for all diagrams to keep them uniform
    \tikzset{every picture/.style={scale=0.7, transform shape}}
    
    % --- Diagram (a) ---
    \subfloat[$gg\to S_0$]{
        \begin{tikzpicture}[baseline=(current bounding box.center)]
            %\node at (-2.2,1.5) {\textbf{(a)}};
            \begin{feynman}
                \vertex (g1) at (-2, 1); \vertex (g2) at (-2, -1);
                \vertex (a) at (0, 1); \vertex (b) at (1.5, 0); \vertex (c) at (0, -1);   
                \vertex (i) at (4, 1); \vertex (j) at (4, -1); \vertex (h) at (3, 0);
                \diagram*{
                    (g1) -- [gluon] (a), (g2) -- [gluon] (c),
                    (a) -- [fermion, edge label= $t$] (b) -- [fermion, edge label= $\bar{t}$] (c) -- [fermion] (a),
                    (b) -- [scalar, thick, edge label = \(S_0\),blue] (h)%,
                   % (h) -- [fermion] (i), (h) -- [anti fermion] (j),
                };
                \filldraw[fill=purple] (b) circle (2.5pt); %\filldraw[fill=purple] (h) circle (2.5pt);
            \end{feynman}
        \end{tikzpicture}
    }
    \hfill
    % --- Diagram (b) ---
    \subfloat[$gg\to h\to S_0h$]{
    \begin{tikzpicture}[baseline=(current bounding box.center)]
            \begin{feynman}
                \vertex (g1) at (-2, 1); \vertex (g2) at (-2, -1);
                \vertex (a) at (0, 1); \vertex (b) at (1.5, 0); \vertex (c) at (0, -1);   
                \vertex (i) at (4, 1); \vertex (j) at (4, -1); \vertex (h) at (3, 0);
                \diagram*{
                    (g1) -- [gluon] (a), (g2) -- [gluon] (c),
                    (a) -- [fermion] (b) -- [fermion] (c) -- [fermion] (a),
                    (b) -- [scalar, edge label = $h$] (h),
                    (h) -- [scalar, edge label = \(S_0\), thick, blue] (i), (h) -- [scalar, edge label' = $h$] (j),
                };
                %\filldraw[fill=purple] (b) circle (2.5pt); 
                \filldraw[fill=blue] (h) circle (2.5pt);
            \end{feynman}
        \end{tikzpicture}
    }
    \hfill
    % --- Diagram (c) ---
    \subfloat[$gg\to S_0 h$ (triangle)]{
    \begin{tikzpicture}[baseline=(current bounding box.center)]
        \begin{feynman}
                \vertex (g1) at (-2, 1); \vertex (g2) at (-2, -1);
                \vertex (a) at (0, 1); \vertex (b) at (1.5, 0); \vertex (c) at (0, -1);   
                \vertex (h) at (3.2, 1) {$h$}; \vertex [blue](k) at (3.2,-1) {\(S_0\)};
                \diagram*{
                    (g1) -- [gluon] (a), (g2) -- [gluon] (c),
                    (a) -- [fermion] (b) -- [fermion] (c) -- [fermion] (a),
                    (b) -- [scalar] (h), (b) -- [scalar, thick, blue] (k),
                };
                \filldraw[fill=purple] (b) circle (2.5pt);
            \end{feynman}
        \end{tikzpicture}
    }
    \hfill
    % --- Diagram (d/e) ---
    \subfloat[$gg\to S_0 h$ (box)]{
    \begin{tikzpicture}[baseline=(current bounding box.center)]
            \begin{feynman}
                \vertex (g1) at (-1.6, 1); \vertex (g2) at (-1.6, -1);
                \vertex (a) at (0, 1); \vertex (b1) at (1.8,1); \vertex (b2) at (1.8,-1); \vertex (c) at (0, -1);   
                \vertex (h) at (3.2, 1) {$h$}; \vertex [blue](k) at (3.2,-1) {\(S_0\)};
                \diagram*{
                    (g1) -- [gluon] (a), (g2) -- [gluon] (c),
                    (a) -- [fermion] (b1) -- [fermion] (b2) -- [fermion] (c) -- [fermion] (a),
                    (b1) -- [scalar] (h), (b2) -- [scalar, thick, blue] (k),
                };
                %\filldraw[fill=purple] (b1) circle (2.5pt); 
                \filldraw[fill=purple] (b2) circle (2.5pt);
            \end{feynman}
        \end{tikzpicture}
    }
    \caption{Representative Feynman diagrams for resonant $S_0$ production through ggF. In (a), $S_0$ is produced singly, whereas in (b),(c) and (d), it is produced in association with a Higgs boson. When the $S_0$ further decays, depending on $M_{0}$, we can get a di-top or di-Higgs final state from (a) and
    a $3h$ or a $t\overline{t}h$ final state from the rest.}
    \label{fig:feynman_diagrams}
\end{figure*}
%%%%%%%%%%%%%%%%%%%%%%%%%%%%%%%%%%%%%%%%%%%%%%%%%%%%%%%%%%%%%%%%

In our framework, the couplings are given by a scalar field $\tilde{\lambda}_i = \lambda_i\epsilon(x)$, where $\epsilon(x)$ is a dimensionless scalar field with non-canonical kinetic term of the form $\frac{1}{2}(\Lambda^2/\epsilon^2)(\partial_\mu \epsilon)^2$. Such terms are typical for moduli fields in string theory and appear when fundamental constants are assumed to vary~\cite{Bekenstein:1982eu}. The current value of the coupling is given by $\lambda_i$. To obtain a canonical scalar field with mass dimension one, we introduce a field redefinition $\epsilon = \exp(S_0/\Lambda_0)$, where $S_0$ is the unique new field in our model, and $\Lambda_0$ is the new physics scale. The only new parameters in the theory are $\Lambda_0$ and the mass of the scalar. Unlike conventional Higgs-portal extensions, where a generic singlet inherits interactions via tunable mixing angles, the 1ST does not have this freedom. Because the production cross-section and decay widths are locked to the same scale $\Lambda_0$, a null result cannot be evaded by tuning a mixing parameter, rendering the framework highly predictive with a phenomenology dominated by the scalar mass relative to the Higgs and top-quark thresholds.  

Depending on its mass, the scalar $S_0$ can decay almost exclusively to gluon pairs or Higgs boson pairs or, above the $t\bar{t}$ threshold, to both Higgs and top quark pairs. We recast the ATLAS di-Higgs and di-top data to set lower bounds on $\Lambda_0$. We demonstrate that the High-Luminosity LHC will probe this scale up to several TeV, providing a test of the dynamical origin of the Higgs sector. Since the singlet nature of $S_0$ forbids SM gauge interference, 1ST also offers probes of the Yukawa variation in rare associated production channels in future colliders. In a following paper \cite{inpreparation} we develop the detailed theoretical framework and generalise it to also consider the possibility of a two-scalar theory (2ST) with two distinct scalar fields, driving independently varying couplings. 

%%%%%%%%%%%%%%%%%%%%%%%%%%%%%%%%%%%%%%%%%%%%%%%%%%%%%%%%%%%%%%%%
\section{Effective field theory description}%\noindent
To obtain the interactions of $S_0$, we insert the couplings on the form $\tilde{\lambda}_i = \lambda_i \exp(S_0/\Lambda_0)$ into the SM Lagrangian and expand to leading order in $1/\Lambda_0$. This leads to the Lagrangian
\begin{align}
\label{eq:1STLag}
\mathcal{L} \supset&\ -\frac{S_0}{\Lambda_0} \left[ \lambda_h \left( v^2 h^2 + v h^3 + \frac{1}{4}h^4 \right) - \frac{\lambda_t}{\sqrt{2}}(v + h)\bar{t}t \right],
\end{align}
where $v$ is the Higgs VEV. The scalar $S_0$ thus couples to the SM sectors through dimension-five operators~\cite{Danielsson:2016nyy,Danielsson:2019ftq}. In 1ST, this single field, $S_0$, governs the variations of both the Higgs self-coupling ($\lambda_h$) and the top Yukawa coupling ($\lambda_t$). This setup produces the correlation between the production and decay modes of the new scalar.

The most general scalar potential involving $S_0$ and the SM Higgs doublet $\Phi$ includes terms of the form $\sum \lambda_n S_0^n (\Phi^\dagger\Phi) + \sum \lambda^{\prime}_n S_0^n$. During electroweak symmetry breaking, the interaction $\lambda_1 S_0 (\Phi^\dagger \Phi)$ induces an off-diagonal $S_0 h$ entry in the scalar mass matrix. 
Because such $S_0 \leftrightarrow h$ mixing is severely restricted by LHC measurements, we assume it is negligible and treat $S_0$ as the physical mass eigenstate. The massive scalar $S_0$ can be assumed to originate from an approximate conformal/scale symmetry, acting as a pseudo-Goldstone boson or dilaton. In principle, the approximate nature of the symmetry can have its origin in quantum effects, giving $S_0$ its mass~\cite{Coradeschi:2013gda}. In such a framework, the electroweak scale emerges naturally when the symmetry breaks at $\Lambda_0$. If this scaling affects all fermions in proportion to their mass (i.e., minimal flavour violation~\cite{DAmbrosio:2002vsn}), it naturally prevents problematic flavour-changing interactions and justifies our focus on the top quark at the collider. This also justifies ensuring $M_{0} < \Lambda_0$, which keeps the effective field theory expansion perturbative and well-behaved. Ultimately, the phenomenology is entirely determined by the mass $M_{0}$ and the single new physics scale $\Lambda_0$. We assume the remaining scalar potential parameters are chosen such that the electroweak vacuum remains stable~\cite{Ghosh:2015apa}.

%%%%%%%%%%%%%%%%%%%%%%%%%%%%%%%%%%%%%%%%%%%%%%%%%%%%%%%%%%%%%%%%
\section{Production and decay of $S_0$}
\noindent
At the LHC, $S_0$ is produced predominantly via gluon-gluon fusion (ggF), mediated by a top-quark loop (see Fig.~\ref{fig:feynman_diagrams}). Because the 1ST framework couples $S_0$ to the SM via the interactions as shown in Eq.~\eqref{eq:1STLag}, a single scale, $\Lambda_0$, determines both the production cross-section and the decay widths. The primary two-body decay modes are di-gluon, di-Higgs and di-top, with partial widths given as:
\begin{align}
\Gamma_{gg} &= \frac{\lambda_t^4\alpha_s^2}{8\pi^3}\left(\frac{v^2}{M_{0}}\right)\left(\frac{v}{\Lambda_0}\right)^2\left|\big[ 1 + (1-\tau_t) f(\tau_t) \big] \right|^2,\nonumber\\
\Gamma_{hh} &=  \frac{\lambda_h^2 }{8\pi } \left(\frac{v^2}{M_{0}}\right)\left(\frac{v}{\Lambda_0}\right)^2\sqrt{1 - \tau_h}, \nonumber\\
\Gamma_{t\bar{t}} &=  \frac{3\lambda_t^2 }{8\pi} M_{0}\left(\frac{v}{\Lambda_0}\right)^2 \left(1 - \tau_t\right)^{3/2}.
\end{align}
The digluon decay is top-loop mediated. Here,  $\tau_i = 4m_i^2/M_{0}^2$ and $f(\tau)$ is the standard scalar loop function that appears in the Higgs decay to a pair of gluons or photons~\cite{Gunion:1989we}. Three- and four-body decays, such as $S_0 \rightarrow hhh$, $S_0\rightarrow ht\bar{t}$ or $S_0 \rightarrow hhhh$, remain kinematically suppressed in the mass range of interest.
%%%%%%%%%%%%%%%%%%%%%%%%%%%%%%%%%%%%%%%%%%%%%%%%%%%%%%%%%%%%%%%%

%%%%%%%%%%%%%%%%%%%%%%%%%%%%%%%%%%%%%%%%%%%%%%%%%%%%%%%%%%%%%%%%
\begin{figure*}
 \captionsetup[subfigure]{labelformat=empty}
\centering
\subfloat{\includegraphics[width=0.47\linewidth]{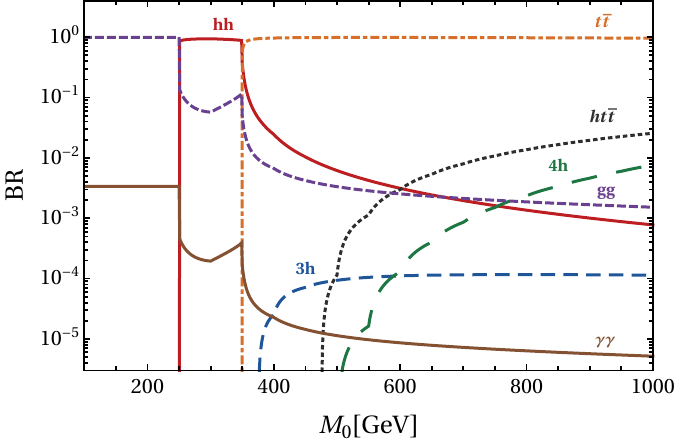}\label{fig:1STBR}}\hfill
\subfloat{\includegraphics[width=0.47\textwidth]{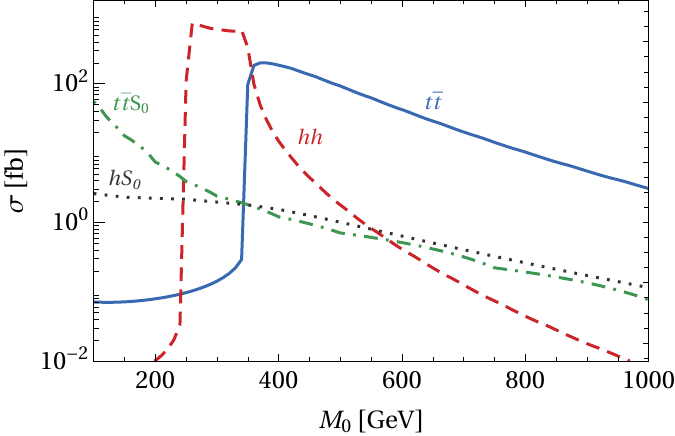}\label{fig:1STXSEC}}
\caption{(Left panel) Branching ratios (BRs) of the scalar $S_0$ as functions of $M_{0}$ in the 1ST model. (Right panel) Cross sections of different production channels of $S_0$. For simplicity, we estimate the $k_{\rm QCD}$ factors from similar Higgs processes. 
\label{fig:BRXSEC}}
%\end{figure*}
%%%%%%%%%%%%%%%%%%%%%%%%%%%%%%%%%%%%%%%%%%%%%%%%%%%%%%%%%%%%%%%%

%%%%%%%%%%%%%%%%%%%%%%%%%%%%%%%%%%%%%%%%%%%%%%%%%%%%%%%%%%%%%%%%
%\begin{figure*}
\centering
\subfloat{\includegraphics[width=0.47\textwidth]{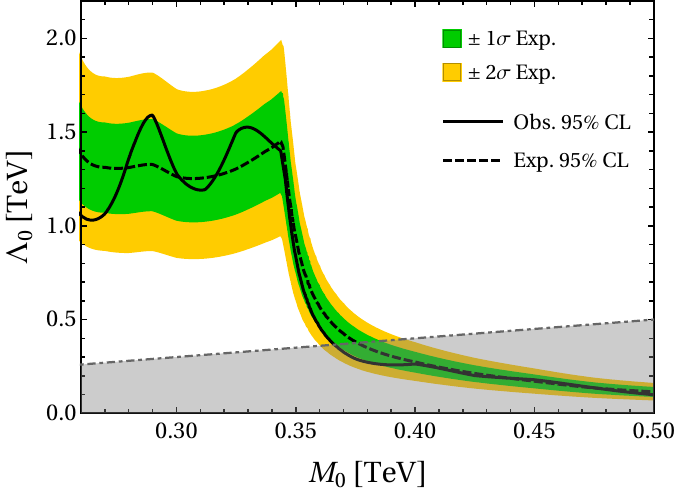}}\hfill
\subfloat{\includegraphics[width=0.47\textwidth]{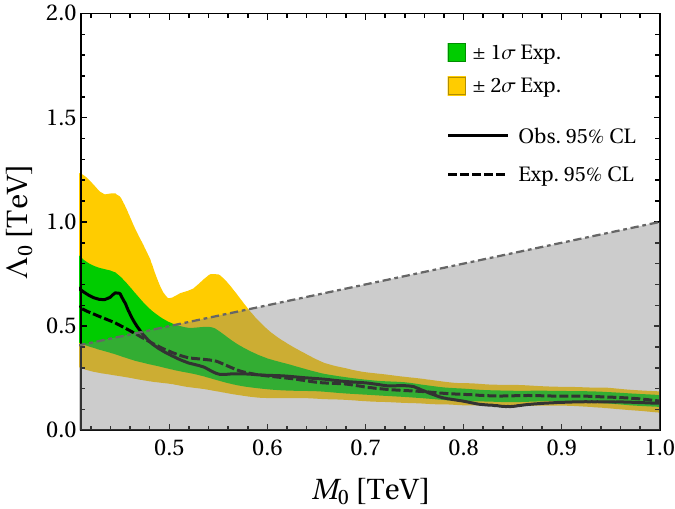}}
\caption{Current 95\% confidence-level lower bounds in the $(M_{0}, \Lambda_0)$ plane. Limits are derived by recasting ATLAS resonance search data in the di-Higgs~\cite{ATLAS:2021ifb} (left panel) and  di-top~\cite{ATLAS:2024vxm} channels (right panel). The regions below the slanted dashed line indicate $M_{0}>\Lambda_0$, where the EFT requires explicit UV completion. }
\label{fig:bounds}
\end{figure*}
%%%%%%%%%%%%%%%%%%%%%%%%%%%%%%%%%%%%%%%%%%%%%%%%%%%%%%%%%%%%%%%%

% %%%%%%%%%%%%%%%%%%%%%%%%%%%%%%%%%%%%%%%%%%%%%%%%%%%%%%%%%%%%%%%%
\begin{figure}[!t]
\centering
\includegraphics[width=\columnwidth]{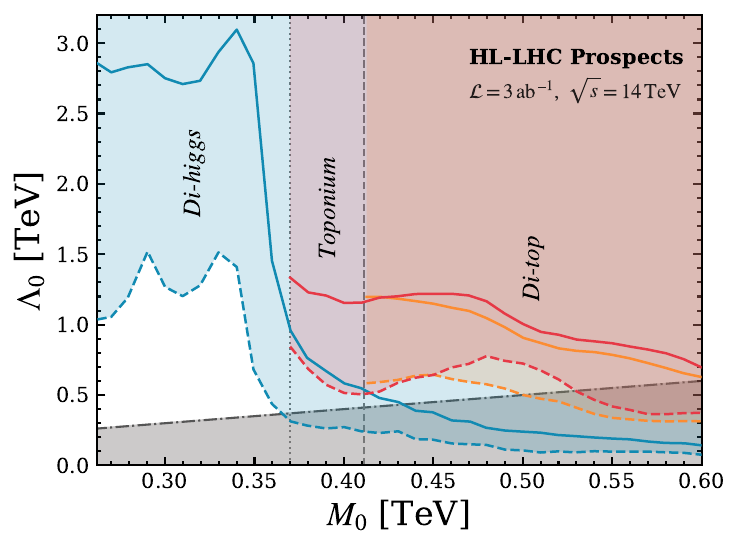}
 \caption{\label{fig:reach} Projected HL-LHC reach in the $(M_{0}, \Lambda_0)$ plane. The shaded blocks show the space left untouched by current LHC searches. Solid lines cutting through these regions show what the HL-LHC can probe through di-Higgs (blue) and di-top (orange) production, while dashed lines mark current limits for comparison. The red curve adds the projection from the recent CMS toponium search~\cite{CMS:2025dzq}. The vertical dashed line marks the $0.4$ TeV lower mass limit of the ATLAS search~\cite{ATLAS:2024vxm}, and the grey area at the bottom shows where $M_{0}>\Lambda_0$ and the effective theory requires a UV completion.}
 \end{figure}
% %%%%%%%%%%%%%%%%%%%%%%%%%%%%%%%%%%%%%%%%%%%%%%%%%%%%%%%%%%%%%%%%

As shown in Fig.~\ref{fig:BRXSEC}, $S_0$ predominantly decays to two gluons below $2m_h$ and almost exclusively to $hh$ in the range $2m_h<M_{0}<2m_t$. Above the top threshold, the $t\bar{t}$ mode dominates, approaching almost $100\%$ branching ratio at higher masses. Because the BRs saturate near unity in their respective regimes, the resonance event rate alone is insufficient to determine $\Lambda_0$. Instead, the coupling can be extracted through interference effects and the total decay width, $\Gamma_{S_0}$. Below $2m_t$, the $\lambda_h(v/\Lambda_0)$ vertex can be probed by the off-shell $S_0$ tail interfering with the non-resonant SM di-Higgs continuum. Above $2m_t$, the $S_0$ resonance interferes with the large SM $gg \to t\bar{t}$ background, producing a characteristic peak-dip lineshape whose phase and magnitude directly map to $\lambda_t(v/\Lambda_0)$.

The shared dependence on $\Lambda_0$ correlates the production rate with the decay signatures, allowing us to set bounds on the $(M_{0},\Lambda_0)$ plane using current LHC di-Higgs and di-top resonance searches. For example, at $M_{0} = 300$ GeV, a scale of $\Lambda_0 = 1.26$ TeV yields a ggF cross-section of approximately $0.4$ pb. We derive the current exclusion limits by recasting ATLAS di-Higgs data~\cite{ATLAS:2021ifb} and extracting upper limits on the coupling modifier from ATLAS di-top searches~\cite{ATLAS:2024vxm}. The resulting bounds on $\Lambda_0$ as a function of mass are shown in Fig.~\ref{fig:bounds}.

The severity of these limits restricts secondary production channels. In principle, the $p p \to t S_0 j$ process offers an alternative, interference-free probe of the top-Yukawa variation. In the SM, the analogous $p p \to t h j$ cross-section is tiny due to destructive gauge interference. Because $S_0$ is a gauge singlet, it avoids this cancellation entirely. However, because the bounds in Fig.~\ref{fig:bounds} force $\Lambda_0 \gtrsim 1$ TeV across the mass range, the overall $(v/\Lambda_0)^2$ penalty suppresses the $p p \to t S_0 j$ cross-section well below observable levels. Similarly, while the associated production $pp \to t\bar{t}S_0$ avoids gauge cancellation and yields a larger cross-section, it is overwhelmed by the massive $t\bar{t}$ plus jets background.

Similarly, even though the tree-level two-body decays dominate the branching ratios, the loop-induced channels $S_0 \to gg$ and $S_0 \to \gamma\gamma$ provide a unique, smoking-gun signature of the 1ST framework. 
Because $S_0$ is a gauge singlet, it lacks tree-level couplings to the $W$ and $Z$ bosons. Consequently, both the digluon and diphoton decays are mediated by the same top-quark loop. The loop factor cancels in the ratio of the decay widths:
\begin{align}
\frac{\Gamma_{\gamma\gamma}}{\Gamma_{gg}} = \frac{128}{256} \left(\frac{\alpha}{\alpha_s}\right)^2 (N_c Q_t^2)^2 = \frac{8}{9}\left(\frac{\alpha}{\alpha_s}\right)^2,    
\end{align}
where the $128/256$ factor arises from the colour trace in the digluon amplitude versus the pure QED phase space of the diphoton amplitude; $N_c=3$ and $Q_t=2/3$ for the top quark. 

Because these loops are suppressed relative to the tree-level di-Higgs mode, their cross-sections are small but cleanly predictable. Similar arguments apply to $S_0\to Z\gamma$ and $S_0\to ZZ$ decays as well, which are further suppressed. For the benchmark $M_{0} = 300$ GeV with a total ggF production cross-section of $0.4$ pb, the observable digluon cross-section ($\sigma \times \text{BR}$) is on the order of a few fb, while the diphoton cross-section lies in the $\mathcal{O}(10^{-3})$ fb regime. While experimentally challenging due to the rate, observing this resonance in the diphoton channel without $W$-boson loop interference (as reflected in the photon-gluon ratio ) would confirm the gauge-singlet nature of the scalar. This could be a vital signature, especially below the di-Higgs threshold, where the gluon decay dominates. The scalar $S_0$ also alters the top-bottom quark mass splitting inside vacuum polarisation loops, implying a non-zero shift in the precision electroweak $T$-parameter. However, the $1/\Lambda_0^2$ suppression ensures this effect remains a secondary, subleading constraint.

%%%%%%%%%%%%%%%%%%%%%%%%%%%%%%%%%%%%%%%%%%%%%%%%%%%%%%%%%%%%%%%%
\section{Multi-Higgs and di-top channels}
\noindent
Correlated multi-Higgs signatures, including di-Higgs and tri-Higgs production, together with resonant di-top production, provide characteristic collider probes of the 1ST model. We briefly discuss these channels below.

\noindent--- \emph{Di-Higgs:} The di-Higgs process, $gg\to S_0\to hh$, receives a significant resonant enhancement once the di-Higgs threshold is crossed. Above the $t\bar t$ threshold, however, the resonant di-top mode becomes dominant due to the opening of the $S_0\to t\bar t$ decay channel. In the intermediate mass range, $2m_h < M_{0} < 2m_t$,
the di-Higgs production cross section can exceed the SM prediction by roughly a factor of $15$ for $\Lambda_0=1$~TeV, rendering the 1ST model highly testable through resonant di-Higgs searches. Ref.~\cite{ATLAS:2025eii} show that a sensitivity of about two times $\sigma_{hh}^{\rm SM}$ can be achieved at the $5\sigma$ level. Pairing the variations of $\lambda_t$ and $\lambda_h$ maximises this di-Higgs rate; no other combination of couplings affects this rate as effectively.

\noindent--- \emph{Tri-Higgs:} In the SM, the tri-Higgs production cross section is just about $0.1$~fb at $14$ TeV LHC~\cite{Abouabid:2024gms,deFlorian:2013jea}, making its observation unrealistic even at the HL-LHC. In contrast, the 1ST model predicts a substantial enhancement. For instance, for $M_{0}=300$~GeV and $\Lambda_0=1$~TeV, the $\sigma_{hhh}^{\rm 1ST}/\sigma_{hhh}^{\rm SM}$ ratio can be as high as $20$. Moreover, two of the three Higgses form a resonance if $M_{0}>2m_h$. Such a sizeable enhancement with a resonance feature could render this channel observable at future LHC runs, thereby providing a striking signature of the 1ST framework.

\noindent--- \emph{Di-top:} In the 1ST model, the resonant di-top production cross section reaches a few hundred fb above the $t\bar t$ threshold. Although this is about three orders of magnitude smaller than the inclusive SM $pp\to t\bar t$ background, the signal originates from resonant production and, therefore, exhibits distinct kinematic features compared to the SM continuum. Advanced analysis strategies, including machine-learning-based techniques, can use these differences to isolate the signal at future collider experiments.

We estimate the projected $2\sigma$ upper bounds on the signal cross-section times branching ratio ($\sigma \times \text{BR}$) by scaling current expected limits with luminosity. Assuming background uncertainties scale statistically, the space the HL-LHC can probe satisfies:
\begin{equation}
\label{eq:HLLHCcondition}
(\sigma \times \text{BR})_\text{exp} \sqrt{\frac{L_{\rm LHC}}{L_{\rm HL\text{-}LHC}}} < (\sigma \times \text{BR})_\text{th} < (\sigma \times \text{BR})_\text{obs},
\end{equation}
where $(\sigma \times \text{BR})_\text{obs}$ and $(\sigma \times \text{BR})_\text{exp}$ are the observed and expected upper limits at the present integrated luminosity $L_{\rm LHC}$, and $L_{\rm HL\text{-}LHC}$ denotes the future target luminosity. The theoretical prediction, $(\sigma \times \text{BR})_\text{th}$, is entirely determined by the mass $M_{0}$ and the fundamental scale $\Lambda_0$. 

Using the current Run~2 luminosity of $L_{\rm LHC} = 140~\mathrm{fb}^{-1}$ and an HL-LHC target of $L_{\rm HL\text{-}LHC} = 3000~\mathrm{fb}^{-1}$, the cross-section limit improves by a factor of $\sqrt{3000/140} \approx 4.6$. Because the 1ST theoretical cross-section scales precisely as $1/\Lambda_0^2$, this translates to roughly a factor of two improvement in the lower bound on $\Lambda_0$.

We show these projections in Fig.~\ref{fig:reach}. The areas below the dashed lines are currently excluded, while the solid lines show the projected HL-LHC reach. The plot shows the kinematic transition in the model: below $2m_t$, the bounds are driven by the di-Higgs channel. Above $2m_t$, the di-Higgs branching ratio drops, and the $t\bar{t}$ resonance search takes over. We also include the recent toponium search from CMS~\cite{CMS:2025dzq}; although toponium is a pseudoscalar, it offers a useful guide since $S_0$ is also spinless. The vertical dashed line marks the $0.4$ TeV lower mass limit of the ATLAS search~\cite{ATLAS:2024vxm}. Together, these two complementary channels will probe $\Lambda_0$ up to $3\text{--}4$~TeV across the mass range.

%%%%%%%%%%%%%%%%%%%%%%%%%%%%%%%%%%%%%%%%%%%%%%%%%%%%%%%%%%%%%%%%
\section{Conclusion}\noindent
In this letter, we introduced the One Scalar Theory, a minimalist framework where a single real singlet scalar dynamically generates both the Higgs self-coupling and the top Yukawa coupling. Unlike generic Higgs-portal extensions, a single fundamental scale, $\Lambda_0$, governs the 1ST. By locking the production cross-section to the decay widths and eliminating tunable scalar mixing, the model becomes exceptionally predictable. We showed that the $2m_t$ kinematic threshold partitions its collider phenomenology, dividing the parameter space into di-Higgs and di-top resonance search regimes. By recasting current ATLAS data, we placed lower bounds on $\Lambda_0$ at the TeV scale and demonstrated that the High-Luminosity LHC will test this framework, extending the reach of $\Lambda_0$ up to $3\text{--}4$~TeV.

The rigid structure of the 1ST offers distinct phenomenological advantages. Because the scalar specifically determines the Yukawa and Higgs self-interactions, it bypasses the standard $WW$ and $ZZ$ gauge decays typical of portal mixing, yielding a unique signature free from massive vector-boson backgrounds. As a direct consequence, its loop-induced diphoton and digluon decays are mediated exclusively by the top quark, locking their partial widths into a rigid, parameter-free ratio. While challenging to observe at current luminosities, this strict proportionality serves as a definitive veto against generic portal theories at future precision colliders. Furthermore, while heavy moduli often face the cosmological moduli problem, the unsuppressed decays of $S_0$ into SM pairs ensure it safely depletes in the early universe without disrupting Big Bang Nucleosynthesis. 

While a complete ultraviolet theory must ultimately address the deeper challenges in conformal frameworks (like the exact mechanism of spontaneous scale breaking, the generation of a stable effective potential, and the cosmological constant problem), this effective theory isolates the testable TeV-scale consequences of this paradigm. Ultimately, the 1ST provides a highly predictive avenue to explore whether the fundamental constants of the electroweak scale are dynamical in origin, with the HL-LHC poised to deliver a definitive verdict on this mechanism.\bigskip

\begin{acknowledgments}
\noindent
We acknowledge support from the Royal Society of Arts and Sciences of Uppsala (KVSU). T.M. acknowledges partial support from the SERB/ANRF, Government of India, through the Core Research Grant No.~CRG/2023/007031.
\end{acknowledgments}

\bibliography{reference}

\begin{thebibliography}{18}%
\makeatletter
\providecommand \@ifxundefined [1]{%
 \@ifx{#1\undefined}
}%
\providecommand \@ifnum [1]{%
 \ifnum #1\expandafter \@firstoftwo
 \else \expandafter \@secondoftwo
 \fi
}%
\providecommand \@ifx [1]{%
 \ifx #1\expandafter \@firstoftwo
 \else \expandafter \@secondoftwo
 \fi
}%
\providecommand \natexlab [1]{#1}%
\providecommand \enquote  [1]{``#1''}%
\providecommand \bibnamefont  [1]{#1}%
\providecommand \bibfnamefont [1]{#1}%
\providecommand \citenamefont [1]{#1}%
\providecommand \href@noop [0]{\@secondoftwo}%
\providecommand \href [0]{\begingroup \@sanitize@url \@href}%
\providecommand \@href[1]{\@@startlink{#1}\@@href}%
\providecommand \@@href[1]{\endgroup#1\@@endlink}%
\providecommand \@sanitize@url [0]{\catcode `\\12\catcode `\$12\catcode
  `\&12\catcode `\#12\catcode `\^12\catcode `\_12\catcode `\%12\relax}%
\providecommand \@@startlink[1]{}%
\providecommand \@@endlink[0]{}%
\providecommand \url  [0]{\begingroup\@sanitize@url \@url }%
\providecommand \@url [1]{\endgroup\@href {#1}{\urlprefix }}%
\providecommand \urlprefix  [0]{URL }%
\providecommand \Eprint [0]{\href }%
\providecommand \doibase [0]{https://doi.org/}%
\providecommand \selectlanguage [0]{\@gobble}%
\providecommand \bibinfo  [0]{\@secondoftwo}%
\providecommand \bibfield  [0]{\@secondoftwo}%
\providecommand \translation [1]{[#1]}%
\providecommand \BibitemOpen [0]{}%
\providecommand \bibitemStop [0]{}%
\providecommand \bibitemNoStop [0]{.\EOS\space}%
\providecommand \EOS [0]{\spacefactor3000\relax}%
\providecommand \BibitemShut  [1]{\csname bibitem#1\endcsname}%
\let\auto@bib@innerbib\@empty
%</preamble>
\bibitem [{\citenamefont {Bekenstein}(1982)}]{Bekenstein:1982eu}%
  \BibitemOpen
  \bibfield  {author} {\bibinfo {author} {\bibfnamefont {J.~D.}\ \bibnamefont
  {Bekenstein}},\ }\bibfield  {title} {\bibinfo {title} {{Fine Structure
  Constant: Is It Really a Constant?}},\ }\href
  {https://doi.org/10.1103/PhysRevD.25.1527} {\bibfield  {journal} {\bibinfo
  {journal} {Phys. Rev. D}\ }\textbf {\bibinfo {volume} {25}},\ \bibinfo
  {pages} {1527} (\bibinfo {year} {1982})}\BibitemShut {NoStop}%
\bibitem [{\citenamefont {Danielsson}\ \emph {et~al.}(2017)\citenamefont
  {Danielsson}, \citenamefont {Enberg}, \citenamefont {Ingelman},\ and\
  \citenamefont {Mandal}}]{Danielsson:2016nyy}%
  \BibitemOpen
  \bibfield  {author} {\bibinfo {author} {\bibfnamefont {U.}~\bibnamefont
  {Danielsson}}, \bibinfo {author} {\bibfnamefont {R.}~\bibnamefont {Enberg}},
  \bibinfo {author} {\bibfnamefont {G.}~\bibnamefont {Ingelman}},\ and\
  \bibinfo {author} {\bibfnamefont {T.}~\bibnamefont {Mandal}},\ }\bibfield
  {title} {\bibinfo {title} {{Heavy photophilic scalar at the LHC from a
  varying electromagnetic coupling}},\ }\href
  {https://doi.org/10.1016/j.nuclphysb.2017.04.003} {\bibfield  {journal}
  {\bibinfo  {journal} {Nucl. Phys. B}\ }\textbf {\bibinfo {volume} {919}},\
  \bibinfo {pages} {569} (\bibinfo {year} {2017})},\ \Eprint
  {https://arxiv.org/abs/1601.00624} {arXiv:1601.00624 [hep-ph]} \BibitemShut
  {NoStop}%
\bibitem [{\citenamefont {Danielsson}\ \emph {et~al.}(2019)\citenamefont
  {Danielsson}, \citenamefont {Enberg}, \citenamefont {Ingelman},\ and\
  \citenamefont {Mandal}}]{Danielsson:2019ftq}%
  \BibitemOpen
  \bibfield  {author} {\bibinfo {author} {\bibfnamefont {U.}~\bibnamefont
  {Danielsson}}, \bibinfo {author} {\bibfnamefont {R.}~\bibnamefont {Enberg}},
  \bibinfo {author} {\bibfnamefont {G.}~\bibnamefont {Ingelman}},\ and\
  \bibinfo {author} {\bibfnamefont {T.}~\bibnamefont {Mandal}},\ }\bibfield
  {title} {\bibinfo {title} {{Varying gauge couplings and collider
  phenomenology}},\ }\href {https://doi.org/10.1103/PhysRevD.100.055028}
  {\bibfield  {journal} {\bibinfo  {journal} {Phys. Rev. D}\ }\textbf {\bibinfo
  {volume} {100}},\ \bibinfo {pages} {055028} (\bibinfo {year} {2019})},\
  \Eprint {https://arxiv.org/abs/1905.11314} {arXiv:1905.11314 [hep-ph]}
  \BibitemShut {NoStop}%
\bibitem [{\citenamefont {Baldes}\ \emph {et~al.}(2018)\citenamefont {Baldes},
  \citenamefont {Konstandin},\ and\ \citenamefont {Servant}}]{Baldes:2016rqn}%
  \BibitemOpen
  \bibfield  {author} {\bibinfo {author} {\bibfnamefont {I.}~\bibnamefont
  {Baldes}}, \bibinfo {author} {\bibfnamefont {T.}~\bibnamefont {Konstandin}},\
  and\ \bibinfo {author} {\bibfnamefont {G.}~\bibnamefont {Servant}},\
  }\bibfield  {title} {\bibinfo {title} {{A first-order electroweak phase
  transition from varying Yukawas}},\ }\href
  {https://doi.org/10.1016/j.physletb.2018.10.015} {\bibfield  {journal}
  {\bibinfo  {journal} {Phys. Lett. B}\ }\textbf {\bibinfo {volume} {786}},\
  \bibinfo {pages} {373} (\bibinfo {year} {2018})},\ \Eprint
  {https://arxiv.org/abs/1604.04526} {arXiv:1604.04526 [hep-ph]} \BibitemShut
  {NoStop}%
\bibitem [{\citenamefont {Braconi}\ \emph {et~al.}(2019)\citenamefont
  {Braconi}, \citenamefont {Chen},\ and\ \citenamefont
  {Gaswint}}]{Braconi:2018gxo}%
  \BibitemOpen
  \bibfield  {author} {\bibinfo {author} {\bibfnamefont {A.}~\bibnamefont
  {Braconi}}, \bibinfo {author} {\bibfnamefont {M.-C.}\ \bibnamefont {Chen}},\
  and\ \bibinfo {author} {\bibfnamefont {G.}~\bibnamefont {Gaswint}},\
  }\bibfield  {title} {\bibinfo {title} {{Revisiting electroweak phase
  transition with varying Yukawa coupling constants}},\ }\href
  {https://doi.org/10.1103/PhysRevD.100.015032} {\bibfield  {journal} {\bibinfo
   {journal} {Phys. Rev. D}\ }\textbf {\bibinfo {volume} {100}},\ \bibinfo
  {pages} {015032} (\bibinfo {year} {2019})},\ \Eprint
  {https://arxiv.org/abs/1810.02522} {arXiv:1810.02522 [hep-ph]} \BibitemShut
  {NoStop}%
\bibitem [{\citenamefont {Ipek}\ and\ \citenamefont
  {Tait}(2019)}]{Ipek:2018lhm}%
  \BibitemOpen
  \bibfield  {author} {\bibinfo {author} {\bibfnamefont {S.}~\bibnamefont
  {Ipek}}\ and\ \bibinfo {author} {\bibfnamefont {T.~M.~P.}\ \bibnamefont
  {Tait}},\ }\bibfield  {title} {\bibinfo {title} {{Early Cosmological Period
  of QCD Confinement}},\ }\href
  {https://doi.org/10.1103/PhysRevLett.122.112001} {\bibfield  {journal}
  {\bibinfo  {journal} {Phys. Rev. Lett.}\ }\textbf {\bibinfo {volume} {122}},\
  \bibinfo {pages} {112001} (\bibinfo {year} {2019})},\ \Eprint
  {https://arxiv.org/abs/1811.00559} {arXiv:1811.00559 [hep-ph]} \BibitemShut
  {NoStop}%
\bibitem [{\citenamefont {Chatrchyan}\ \emph {et~al.}(2026)\citenamefont
  {Chatrchyan}, \citenamefont {Marsh},\ and\ \citenamefont
  {Nikolis}}]{Chatrchyan:2025wop}%
  \BibitemOpen
  \bibfield  {author} {\bibinfo {author} {\bibfnamefont {A.}~\bibnamefont
  {Chatrchyan}}, \bibinfo {author} {\bibfnamefont {M.~C.~D.}\ \bibnamefont
  {Marsh}},\ and\ \bibinfo {author} {\bibfnamefont {C.}~\bibnamefont
  {Nikolis}},\ }\bibfield  {title} {\bibinfo {title} {{Gravitational Waves from
  a Dilaton-Induced, First-Order QCD Phase Transition}},\ }\href
  {https://doi.org/10.1103/2v2f-1jvz} {\bibfield  {journal} {\bibinfo
  {journal} {Phys. Rev. Lett.}\ }\textbf {\bibinfo {volume} {136}},\ \bibinfo
  {pages} {041005} (\bibinfo {year} {2026})},\ \Eprint
  {https://arxiv.org/abs/2507.01191} {arXiv:2507.01191 [hep-ph]} \BibitemShut
  {NoStop}%
\bibitem [{\citenamefont {Danielsson}\ \emph {et~al.}(2026)\citenamefont
  {Danielsson}, \citenamefont {Enberg}, \citenamefont {Ingelman}, \citenamefont
  {Kundu}, \citenamefont {Mandal},\ and\ \citenamefont
  {Mitra}}]{inpreparation}%
  \BibitemOpen
  \bibfield  {author} {\bibinfo {author} {\bibfnamefont {U.}~\bibnamefont
  {Danielsson}}, \bibinfo {author} {\bibfnamefont {R.}~\bibnamefont {Enberg}},
  \bibinfo {author} {\bibfnamefont {G.}~\bibnamefont {Ingelman}}, \bibinfo
  {author} {\bibfnamefont {S.}~\bibnamefont {Kundu}}, \bibinfo {author}
  {\bibfnamefont {T.}~\bibnamefont {Mandal}},\ and\ \bibinfo {author}
  {\bibfnamefont {S.}~\bibnamefont {Mitra}},\ }\bibfield  {title} {\bibinfo
  {title} {Multi-{Higgs} production from a theory of varying {Higgs} quartic
  and top {Yukawa} couplings}} (\bibinfo {year} {2026}),\ \bibinfo {note}
  {manuscript in preparation}\BibitemShut {NoStop}%
\bibitem [{\citenamefont {Coradeschi}\ \emph {et~al.}(2013)\citenamefont
  {Coradeschi}, \citenamefont {Lodone}, \citenamefont {Pappadopulo},
  \citenamefont {Rattazzi},\ and\ \citenamefont {Vitale}}]{Coradeschi:2013gda}%
  \BibitemOpen
  \bibfield  {author} {\bibinfo {author} {\bibfnamefont {F.}~\bibnamefont
  {Coradeschi}}, \bibinfo {author} {\bibfnamefont {P.}~\bibnamefont {Lodone}},
  \bibinfo {author} {\bibfnamefont {D.}~\bibnamefont {Pappadopulo}}, \bibinfo
  {author} {\bibfnamefont {R.}~\bibnamefont {Rattazzi}},\ and\ \bibinfo
  {author} {\bibfnamefont {L.}~\bibnamefont {Vitale}},\ }\bibfield  {title}
  {\bibinfo {title} {{A naturally light dilaton}},\ }\href
  {https://doi.org/10.1007/JHEP11(2013)057} {\bibfield  {journal} {\bibinfo
  {journal} {JHEP}\ }\textbf {\bibinfo {volume} {11}},\ \bibinfo {pages}
  {057}},\ \Eprint {https://arxiv.org/abs/1306.4601} {arXiv:1306.4601 [hep-th]}
  \BibitemShut {NoStop}%
\bibitem [{\citenamefont {D'Ambrosio}\ \emph {et~al.}(2002)\citenamefont
  {D'Ambrosio}, \citenamefont {Giudice}, \citenamefont {Isidori},\ and\
  \citenamefont {Strumia}}]{DAmbrosio:2002vsn}%
  \BibitemOpen
  \bibfield  {author} {\bibinfo {author} {\bibfnamefont {G.}~\bibnamefont
  {D'Ambrosio}}, \bibinfo {author} {\bibfnamefont {G.~F.}\ \bibnamefont
  {Giudice}}, \bibinfo {author} {\bibfnamefont {G.}~\bibnamefont {Isidori}},\
  and\ \bibinfo {author} {\bibfnamefont {A.}~\bibnamefont {Strumia}},\
  }\bibfield  {title} {\bibinfo {title} {{Minimal flavor violation: An
  Effective field theory approach}},\ }\href
  {https://doi.org/10.1016/S0550-3213(02)00836-2} {\bibfield  {journal}
  {\bibinfo  {journal} {Nucl. Phys. B}\ }\textbf {\bibinfo {volume} {645}},\
  \bibinfo {pages} {155} (\bibinfo {year} {2002})},\ \Eprint
  {https://arxiv.org/abs/hep-ph/0207036} {arXiv:hep-ph/0207036} \BibitemShut
  {NoStop}%
\bibitem [{\citenamefont {Ghosh}\ \emph {et~al.}(2016)\citenamefont {Ghosh},
  \citenamefont {Kundu},\ and\ \citenamefont {Ray}}]{Ghosh:2015apa}%
  \BibitemOpen
  \bibfield  {author} {\bibinfo {author} {\bibfnamefont {S.}~\bibnamefont
  {Ghosh}}, \bibinfo {author} {\bibfnamefont {A.}~\bibnamefont {Kundu}},\ and\
  \bibinfo {author} {\bibfnamefont {S.}~\bibnamefont {Ray}},\ }\bibfield
  {title} {\bibinfo {title} {{Potential of a singlet scalar enhanced Standard
  Model}},\ }\href {https://doi.org/10.1103/PhysRevD.93.115034} {\bibfield
  {journal} {\bibinfo  {journal} {Phys. Rev. D}\ }\textbf {\bibinfo {volume}
  {93}},\ \bibinfo {pages} {115034} (\bibinfo {year} {2016})},\ \Eprint
  {https://arxiv.org/abs/1512.05786} {arXiv:1512.05786 [hep-ph]} \BibitemShut
  {NoStop}%
\bibitem [{\citenamefont {Gunion}\ \emph {et~al.}(2000)\citenamefont {Gunion},
  \citenamefont {Haber}, \citenamefont {Kane},\ and\ \citenamefont
  {Dawson}}]{Gunion:1989we}%
  \BibitemOpen
  \bibfield  {author} {\bibinfo {author} {\bibfnamefont {J.~F.}\ \bibnamefont
  {Gunion}}, \bibinfo {author} {\bibfnamefont {H.~E.}\ \bibnamefont {Haber}},
  \bibinfo {author} {\bibfnamefont {G.~L.}\ \bibnamefont {Kane}},\ and\
  \bibinfo {author} {\bibfnamefont {S.}~\bibnamefont {Dawson}},\ }\href
  {https://doi.org/10.1201/9780429496448} {\emph {\bibinfo {title} {{The Higgs
  Hunter's Guide}}}},\ Vol.~\bibinfo {volume} {80}\ (\bibinfo {year}
  {2000})\BibitemShut {NoStop}%
\bibitem [{\citenamefont {Aad}\ \emph {et~al.}(2022)\citenamefont {Aad} \emph
  {et~al.}}]{ATLAS:2021ifb}%
  \BibitemOpen
  \bibfield  {author} {\bibinfo {author} {\bibfnamefont {G.}~\bibnamefont
  {Aad}} \emph {et~al.} (\bibinfo {collaboration} {ATLAS}),\ }\bibfield
  {title} {\bibinfo {title} {{Search for Higgs boson pair production in the two
  bottom quarks plus two photons final state in $pp$ collisions at
  $\sqrt{s}=13$ TeV with the ATLAS detector}},\ }\href
  {https://doi.org/10.1103/PhysRevD.106.052001} {\bibfield  {journal} {\bibinfo
   {journal} {Phys. Rev. D}\ }\textbf {\bibinfo {volume} {106}},\ \bibinfo
  {pages} {052001} (\bibinfo {year} {2022})},\ \Eprint
  {https://arxiv.org/abs/2112.11876} {arXiv:2112.11876 [hep-ex]} \BibitemShut
  {NoStop}%
\bibitem [{\citenamefont {Aad}\ \emph {et~al.}(2024)\citenamefont {Aad} \emph
  {et~al.}}]{ATLAS:2024vxm}%
  \BibitemOpen
  \bibfield  {author} {\bibinfo {author} {\bibfnamefont {G.}~\bibnamefont
  {Aad}} \emph {et~al.} (\bibinfo {collaboration} {ATLAS}),\ }\bibfield
  {title} {\bibinfo {title} {{Search for heavy neutral Higgs bosons decaying
  into a top quark pair in 140 fb$^{-1}$ of proton-proton collision data at $
  \sqrt{s} $ = 13 TeV with the ATLAS detector}},\ }\href
  {https://doi.org/10.1007/JHEP08(2024)013} {\bibfield  {journal} {\bibinfo
  {journal} {JHEP}\ }\textbf {\bibinfo {volume} {08}},\ \bibinfo {pages}
  {013}},\ \Eprint {https://arxiv.org/abs/2404.18986} {arXiv:2404.18986
  [hep-ex]} \BibitemShut {NoStop}%
\bibitem [{\citenamefont {Hayrapetyan}\ \emph {et~al.}(2025)\citenamefont
  {Hayrapetyan} \emph {et~al.}}]{CMS:2025dzq}%
  \BibitemOpen
  \bibfield  {author} {\bibinfo {author} {\bibfnamefont {A.}~\bibnamefont
  {Hayrapetyan}} \emph {et~al.} (\bibinfo {collaboration} {CMS}),\ }\bibfield
  {title} {\bibinfo {title} {{Search for heavy pseudoscalar and scalar bosons
  decaying to a top quark pair in proton{\textendash}proton collisions at
  $\sqrt{s} = 13\,\textrm{TeV}$}},\ }\href
  {https://doi.org/10.1088/1361-6633/ae2207} {\bibfield  {journal} {\bibinfo
  {journal} {Rept. Prog. Phys.}\ }\textbf {\bibinfo {volume} {88}},\ \bibinfo
  {pages} {127801} (\bibinfo {year} {2025})},\ \Eprint
  {https://arxiv.org/abs/2507.05119} {arXiv:2507.05119 [hep-ex]} \BibitemShut
  {NoStop}%
\bibitem [{\citenamefont {Aad}\ \emph {et~al.}(2025)\citenamefont {Aad} \emph
  {et~al.}}]{ATLAS:2025eii}%
  \BibitemOpen
  \bibfield  {author} {\bibinfo {author} {\bibfnamefont {G.}~\bibnamefont
  {Aad}} \emph {et~al.} (\bibinfo {collaboration} {ATLAS, CMS}),\ }\href@noop
  {} {\bibinfo {title} {{Highlights of the HL-LHC physics projections by ATLAS
  and CMS}}} (\bibinfo {year} {2025}),\ \Eprint
  {https://arxiv.org/abs/2504.00672} {arXiv:2504.00672 [hep-ex]} \BibitemShut
  {NoStop}%
\bibitem [{\citenamefont {Abouabid}\ \emph {et~al.}(2024)\citenamefont
  {Abouabid} \emph {et~al.}}]{Abouabid:2024gms}%
  \BibitemOpen
  \bibfield  {author} {\bibinfo {author} {\bibfnamefont {H.}~\bibnamefont
  {Abouabid}} \emph {et~al.},\ }\bibfield  {title} {\bibinfo {title} {{HHH
  whitepaper}},\ }\href {https://doi.org/10.1140/epjc/s10052-024-13376-3}
  {\bibfield  {journal} {\bibinfo  {journal} {Eur. Phys. J. C}\ }\textbf
  {\bibinfo {volume} {84}},\ \bibinfo {pages} {1183} (\bibinfo {year}
  {2024})},\ \Eprint {https://arxiv.org/abs/2407.03015} {arXiv:2407.03015
  [hep-ph]} \BibitemShut {NoStop}%
\bibitem [{\citenamefont {de~Florian}\ and\ \citenamefont
  {Mazzitelli}(2013)}]{deFlorian:2013jea}%
  \BibitemOpen
  \bibfield  {author} {\bibinfo {author} {\bibfnamefont {D.}~\bibnamefont
  {de~Florian}}\ and\ \bibinfo {author} {\bibfnamefont {J.}~\bibnamefont
  {Mazzitelli}},\ }\bibfield  {title} {\bibinfo {title} {{Higgs Boson Pair
  Production at Next-to-Next-to-Leading Order in QCD}},\ }\href
  {https://doi.org/10.1103/PhysRevLett.111.201801} {\bibfield  {journal}
  {\bibinfo  {journal} {Phys. Rev. Lett.}\ }\textbf {\bibinfo {volume} {111}},\
  \bibinfo {pages} {201801} (\bibinfo {year} {2013})},\ \Eprint
  {https://arxiv.org/abs/1309.6594} {arXiv:1309.6594 [hep-ph]} \BibitemShut
  {NoStop}%
\end{thebibliography}%

\end{document}